\documentclass[conference]{IEEEtran}

\usepackage{algorithmic}
\usepackage{amsfonts}
\usepackage{amsmath}
\usepackage{amssymb}
\usepackage{array}
\usepackage{bm}
\usepackage{booktabs}
\usepackage{cite}
\usepackage{colortbl}
\usepackage{comment}
\usepackage{epsfig}
\usepackage{float}
\usepackage{graphicx}
\usepackage{multicol}
\usepackage{multirow}
\usepackage{soul}
\usepackage{subcaption}
\usepackage{textcomp}
\usepackage{url}
\usepackage{xcolor}

\usepackage[hidelinks]{hyperref}

\newcommand*{\CtoXB}{$\text{\textbf{C2x}}_\mathbf{\mathcal{B}}$}
\newcommand*{\CtoXM}{$\text{\textbf{C2x}}_\mathbf{\mathcal{M}}$}

\IEEEoverridecommandlockouts

\begin{document}

\title{Defending Network Intrusion Detection Systems Based on Graph Neural Networks Against Structural Adversarial Attacks%
\thanks{%
© 2025 IEEE. This is the author's accepted manuscript.
The Version of Record is available at
\url{https://doi.org/10.1109/NCA67271.2025.00043}.
This accepted manuscript is licensed under CC BY 4.0.
}}

\author{
\IEEEauthorblockN{
Dimitri Galli\IEEEauthorrefmark{1},
Andrea Venturi\IEEEauthorrefmark{2},
Dario Stabili\IEEEauthorrefmark{1},
Mauro Andreolini\IEEEauthorrefmark{1},
Mirco Marchetti\IEEEauthorrefmark{1}
}
\IEEEauthorblockA{
\IEEEauthorrefmark{1}\textit{Department of Engineering ``Enzo Ferrari''}, \textit{University of Modena and Reggio Emilia}, Modena, Italy\\
\{dimitri.galli, dario.stabili, mauro.andreolini, mirco.marchetti\}@unimore.it
}
\IEEEauthorblockA{
\IEEEauthorrefmark{2}\textit{Department of Digital Security}, \textit{Vicomtech}, San Sebastián, Spain\\
aventuri@vicomtech.org
}
}

\maketitle

\begin{abstract}
Graph Neural Networks (GNNs) represent a promising solution for Machine Learning (ML) based Network Intrusion Detection Systems (NIDS), thanks to their ability to leverage both network flow features and topological patterns. While GNN classifiers demonstrate superior robustness against feature-based adversarial attacks compared to other ML detectors, they remain vulnerable to structural adversarial attacks, where an attacker perturbs the underlying network graph topology by injecting edges or inserting nodes.
Such attacks pose a realistic and severe threat, undermining the reliability of GNN-based NIDS in practical deployments. While countermeasures have been proposed in the literature, they often rely on assumptions that are unrealistic in real-world cybersecurity scenarios.
In this paper, we propose a defense framework based on adversarial training to strengthen GNN-based NIDS against structural attacks. We generate adversarial samples by strategically replacing the source and destination nodes in benign network flows, thereby efficiently mimicking edge injection attacks. We evaluate our approach on two widely used datasets (CTU-13 and TON-IoT) using E-GraphSAGE as the base GNN classifier. Experimental results show that our approach produces hardened detectors with superior detection performance on clean graphs and enhanced robustness against structural adversarial attacks.
\end{abstract}
\section{Introduction}
\label{sec:introduction}
The exponential growth of network-based cyberattacks has increased the demand for accurate and adaptive Network Intrusion Detection Systems (NIDS)~\cite{sommer2010outside,apruzzese2023role}.
In recent years, Machine Learning (ML) has emerged as a powerful paradigm for building these systems, enabling the automatic discovery of malicious patterns in network traffic and improving the detection of unknown attacks~\cite{apruzzese2023sok}.

Despite these advances, ML-based NIDS remain highly vulnerable to \textit{adversarial ML attacks}~\cite{biggio2018wild}.
Such attacks carefully perturb input features to cause misclassifications, enabling malicious traffic to evade detection mechanisms.
These manipulations provide attackers with a powerful means to conceal their activities while preserving the intended impact of their attacks~\cite{pierazzi2020intriguing}.
The effectiveness of these attacks represents a concrete threat to the reliable deployment of ML-based NIDS in real-world environments.

Graph Neural Networks (GNNs) have emerged as a promising alternative for ML-based NIDS~\cite{jiang2022graph} due to the inherent graph-like structure of network traffic.
GNN-based NIDS take network traffic in graph form as input data, modeling internal hosts as nodes and communication flows between them as edges.
By analyzing both flow-level information (e.g., duration, byte count) and network topology, GNN-based NIDS have demonstrated superior detection performance compared to traditional ML detector~\cite{lo2022graphsage, venturi2023arganids}.
Empirical studies also show that GNNs exhibit stronger robustness against feature-based adversarial attacks due to their reliance on structural attributes~\cite{pujol2022unveiling}.

However, incorporating network topology in the decision process can expose classifiers to a new class of threats known as \textit{structural adversarial attacks}.
These attacks manipulate the graph topology to induce misclassifications~\cite{zugner2018adversarial}.
Since GNNs heavily rely on neighborhood structures, even small perturbations can dramatically degrade their detection performance.
This is particularly concerning in cybersecurity scenarios, as such attacks can be launched by compromised hosts generating seemingly legitimate, and therefore hard-to-detect, network traffic~\cite{venturi2024problem}.
While the scientific literature has proposed defenses against structural adversarial attacks, only a few have been explored in the cybersecurity domain.
Moreover, many existing methods depend on prior knowledge of attack techniques or require computationally expensive resources, resulting in high latency and limiting their real-time deployment~\cite{venturi2023practical, zhan2025real}.

Among traditional adversarial ML defenses, \textit{adversarial training} stands out as a promising approach to enhancing classifier resilience by retraining ML models on an augmented training set that includes carefully crafted adversarial examples~\cite{goodfellow2014explaining}.
Nevertheless, its effectiveness in the specific context of GNN-based NIDS remains largely unexplored in prior literature.
This work aims to fill this gap by introducing a new defensive method based on adversarial training.
The key contribution lies in a novel strategy for generating adversarial examples, which considers the unique characteristics of the network intrusion detection domain and can be used to harden GNN-based NIDS against structural adversarial attacks.
Specifically, we propose to generate synthetic adversarial flows by modifying both the source and destination endpoints of benign netflows with \textit{low-degree nodes}. These nodes with few connections are structurally more vulnerable and therefore attractive to real-world attackers seeking to evade detection.
Each modified record corresponds to a newly injected benign edge connecting low-degree nodes, thereby perturbing the graph topology.

We apply our approach to an extensive experimental campaign using two popular datasets of enterprise network flows.
Our method targets various types of structural adversarial attacks, including those involving edge additions, node additions, or combinations of both.
The results demonstrate the effectiveness of the proposed strategy, with absolute performance improvements of up to 90\% in the best case against structural perturbations.
Moreover, the hardened classifiers maintain high detection performance even in the absence of adversarial attacks.

The remainder of this paper is organized as follows.
Section~\ref{sec:backgroundrelatedwork} provides background on ML-based NIDS, GNN-based NIDS, as well as adversarial ML attacks and corresponding defense mechanisms.
Section~\ref{sec:threatmodel} defines the threat model.
Section~\ref{sec:defensestrategy} describes the proposed defense strategy.
Section~\ref{sec:experiments} details the testbed.
Section~\ref{sec:results} presents and discusses the experimental results.
Section~\ref{sec:conclusion} concludes the paper with final remarks.
\section{Background and Related Work}
\label{sec:backgroundrelatedwork}
This section provides background on NIDS based on ML and GNNs.
We then motivate our work by examining the threat posed by adversarial attacks and identifying critical gaps in the adversarial defense literature.

\subsection{ML-based NIDS}
\label{subsec:mlnids}
ML-based NIDS analyze and classify network data to detect malicious operations, either by processing raw packets directly~\cite{mirsky2018kitsune} or by summarizing traffic into network flows\footnote{\url{https://www.cisco.com/c/en/us/products/ios-nx-os-software/ios-netflow/}} (or \textit{netflows}) using selected metrics and statistics~\cite{yehezkel2021network}.
Depending on the approach, these systems may rely on labeled data for supervised training or operate in an unsupervised manner, learning normal behavior patterns without explicit labels~\cite{apruzzese2023sok}.

Traditional ML classifiers often consider each sample in isolation.
While this enables rapid processing and the real-time generation of security alerts, analyzing records individually limits the possibility of detecting complex multi-step attacks, which require capturing relational dependencies across multiple connections and hosts~\cite{ahmad2021network,pujol2022unveiling}.

\subsection{GNN-based NIDS}
\label{subsec:gnnnids}
Graph Neural Networks~\cite{scarselli2008graph} have recently emerged as a compelling alternative for cyber threat detection, as they can naturally represent network communication patterns as graph structures~\cite{jiang2022graph}.
GNNs are deep learning models specifically designed to handle graph-structured data, leveraging the premise that cyberattacks exhibit distinctive topological patterns detectable through network structure analysis.

Early research has shown that GNN-based NIDS outperform traditional ML detectors both in adversarial environments~\cite{zhou2021hierarchical, pujol2022unveiling} and in attack-free settings~\cite{lo2022graphsage, venturi2023arganids}.
This performance stems from the GNN's ability to simultaneously process netflow attributes and structural features, resulting in more robust detection capabilities~\cite{zhong2024survey}.

The operational workflow of a GNN-based NIDS involves three key phases.
The first phase builds the \textit{graph} from netflow data.
The most common representation, known as a \textit{flow graph}, maps each unique endpoint to a graph node, while individual netflows are represented as edges connecting the corresponding vertices.
This graph representation allows the detection problem to be formulated as an edge classification task~\cite{lo2022graphsage}.
However, the majority of GNN models are designed for node-level tasks, so alternative representations have been studied to analyze node-associated features during training.
For instance, a \textit{line graph} maps individual netflows directly to nodes and connects them when they share a common host in the corresponding records~\cite{venturi2023arganids}.
Once netflows are transformed into an appropriate graph structure, the GNN is trained to generate \textit{node embeddings} that capture higher-order dependencies through message passing and neighborhood aggregation.
Such embeddings encode both netflow information and network topology in a low-dimensional latent space.
\textit{Edge embeddings} can be derived by concatenating the latent vectors of the respective source and destination nodes~\cite{lo2022graphsage}.
To perform detection, these embeddings are fed into a \textit{classifier} that distinguishes benign from malicious network traffic.

\subsection{Adversarial Attacks}
\label{subsec:adversarialattacks}
Adversarial ML attacks exploit a classifier’s sensitivity to carefully crafted perturbations that mislead the model, causing it to produce incorrect outputs that benefit the attacker~\cite{biggio2013evasion, corona2013adversarial}.
These attacks can occur during the training phase, as in \textit{poisoning attacks}, or inference, as in \textit{evasion attacks}, depending on whether the attacker aims to compromise the ML model before or after deployment.

In network intrusion detection scenarios, adversaries typically manipulate common network traffic attributes (e.g., duration, byte count, packet size) to deceive the ML detector and induce misclassifications~\cite{apruzzese2018evading}.
As GNN-based NIDS rely on both flow features and structural patterns, attackers can evade detection by targeting either netflow information (\textit{feature attacks}) or the graph topology (\textit{structural attacks})~\cite{zugner2018adversarial}.

On the one hand, several studies have demonstrated that GNN-based NIDS are relatively robust to feature attacks due to their reliance on neighborhood structures~\cite{zhou2021hierarchical, pujol2022unveiling, wang2022threatrace, venturi2024problem}.
As a result, feature perturbations pose a comparatively lower risk to these systems.
On the other hand, structural attacks targeting the graph topology represent a greater threat to GNN-based NIDS.
These attacks can severely disrupt GNN performance, with minimal perturbations sometimes enabling near-complete evasion while affecting only a small fraction of the original graph~\cite{zugner2018adversarial, hussain2021structack, venturi2024problem}.

\subsection{Adversarial Defenses}
\label{subsec:adversarialdefenses}
Although various countermeasure strategies have been proposed to defend GNNs against such attacks, only a few have been specifically designed or evaluated in the context of network intrusion detection.
This scarcity of practical defenses highlights a critical gap, as most existing solutions target general graph learning tasks and may not account for the unique characteristics and constraints of NIDS environments.

\textit{Preprocessing-based defenses}~\cite{wu2019adversarial, entezari2020all} sanitize the graph by detecting and removing adversarial manipulations before the input is submitted to the GNN.
While effective against specific structural perturbations (e.g., \textit{edge deletion attacks}~\cite{jin2020graph}), such defenses often depend on prior knowledge of attack strategies, which is rarely available in real-world networks.
These defenses may also inadvertently alter or remove legitimate graph components (e.g., benign edges) that are essential for modeling the network topology.
This can obscure important structural relationships, making it harder for the NIDS to detect both known and novel attack patterns.

\textit{Model-based defenses} modify GNN architectures to improve their resilience to perturbations.
Defenses such as \textit{defensive aggregation}~\cite{zhang2020gnnguard} or \textit{robust message passing}~\cite{zhu2019robust} are designed to mitigate the effect of adversarial modifications on model predictions.
However, these methods typically introduce significant computational overhead, resulting in excessive latency that can delay the detection of malicious activity.
Moreover, sophisticated adaptive attacks can still exploit specialized architectures, highlighting the persistent arms race between attackers and defenders.

\textit{Training-based defenses}~\cite{xu2019topology, xu2020towards} enhance GNN robustness against evasion attacks by leveraging specialized training strategies.
In particular, \textit{adversarial training}~\cite{gosch2023adversarial} aims to harden the classifier by incorporating carefully perturbed samples into the original training set, effectively reshaping the model’s decision boundary to enable accurate classification of adversarial patterns.
However, adversarial training faces several challenges when applied to network intrusion detection tasks.
Compared to other domains, adversarial samples are bound by strict validity requirements that preserve protocol compliance, feature consistency, and temporal dependencies~\cite{pierazzi2020intriguing}.
Indeed, adversarial training is generally effective only against predictable attacks that resemble the generated examples, which can lead to a false sense of robustness~\cite{apruzzese2020deep}.
Although it can enhance robustness against specific threats, it often reduces performance on clean data, thereby limiting its practicality in real-world scenarios~\cite{venturi2024hardening}.

Our work builds on adversarial training to strengthen GNN-based NIDS against structural attacks.
Unlike model-based defenses, it directly operates on the netflows in the training set without altering the classifier architecture.
Notably, we generate adversarial examples by modifying the topology information of benign records, without removing legitimate components.
Since we apply perturbations at the structural level, we preserve original flow features, ensuring that all network constraints are maintained.
\section{Threat Model}
\label{sec:threatmodel}
In this section, we define the \textit{threat model} for this work.
More specifically, we outline the goals, knowledge, and capabilities of both the attacker (Section~\ref{subsec:attackermodel}) and the defender (Section~\ref{subsec:defendermodel}).

\begin{figure}[t]
\centering
\includegraphics[width=\columnwidth]{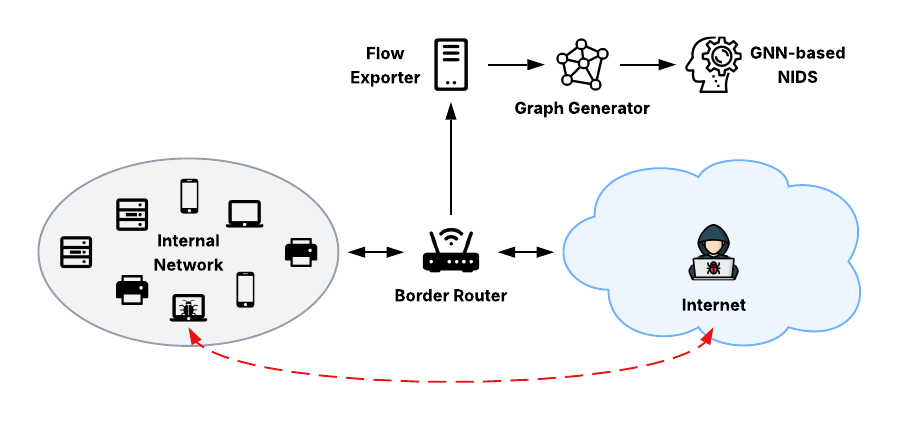}
\caption{Threat model.}
\label{fig:threatmodel}
\end{figure}

Building upon prior work~\cite{apruzzese2022modeling}, we consider a traditional network intrusion detection scenario, as depicted in Fig.~\ref{fig:threatmodel}.
A medium-sized network connects to the Internet through a \textit{border router}, which handles all incoming and outgoing traffic for internal hosts.
The border router also forwards network packets to a \textit{flow exporter}, which collects and extracts flow records.
Once the netflows are produced, a \textit{graph generator} builds a graph representation, which is then used as input to the GNN classifier.
The GNN-based NIDS consists of an \textit{ensemble} of $M$ graph-based individual models.
Each model is trained to identify suspicious activities corresponding to one of the $M$ distinct malicious classes.

\subsection{Attacker Model}
\label{subsec:attackermodel}
The attacker model employed to evaluate the effectiveness of our proposed defense reflects a realistic NIDS threat scenario and builds upon prior work on adversarial attacks in the problem space~\cite{pierazzi2020intriguing, venturi2024problem}.
We consider an attacker located outside the monitored network who has already compromised one host within the enterprise network via common attack vectors, such as software vulnerabilities or weak authentication (see red arrow in Fig.~\ref{fig:threatmodel}).

The attacker’s primary \textit{goal} is to remain undetected while carrying out malicious operations within the organization’s network.
The attacker seeks to evade detection by inducing the NIDS to misclassify malicious communications, all while ensuring that the underlying attacks retain their full effectiveness.

The attacker operates under a gray-box setting with limited \textit{knowledge} of the target system.
In particular, the attacker assumes that a GNN-based NIDS is deployed but lacks detailed information about the classifier configuration, feature set, and graph representation.
The attacker has no access to the training data and operates only at inference time.
Notably, the attacker cannot observe the predictions of the targeted GNN detectors.

The attacker’s \textit{capabilities} are defined by the actions they can perform on the network itself.
While the attacker has complete control over a limited number of compromised hosts, they cannot directly interact with core infrastructure components such as the flow exporter, graph generator, or NIDS.
The attacker cannot manipulate the features of the graph representation submitted to the NIDS (i.e., no feature space perturbations).
Their capabilities are therefore restricted to the following threats:
\begin{itemize}
\item \textit{\CtoXB~attacks}: injecting \textit{benign} edges (denoted by $\mathbf{\mathcal{B}}$) into the graph by establishing legitimate communications from \textit{compromised} hosts (denoted by \textbf{C}) to \textit{random} ones (denoted by \textit{x}) within the network;
\item \textit{\CtoXM~attacks}: injecting \textit{malicious} edges (denoted by $\mathbf{\mathcal{M}}$) into the graph by initiating unauthorized communications from \textit{compromised} hosts (denoted by \textbf{C}) to \textit{random} ones (denoted by \textit{x}) within the network;
\item \textit{\textbf{Add node}~attacks}: injecting \textit{new} nodes into the graph by communicating with external devices not previously part of the network or by compromising internal hosts.
\end{itemize}

\subsection{Defender Model}
\label{subsec:defendermodel}
We consider the defender to be the network administrator responsible for deploying, maintaining, and monitoring the NIDS within the organization, or alternatively, the security vendor that develops and tunes it.
The defender's primary \textit{goal} is to enhance the resilience of GNN-based NIDS against adversarial attacks based on structural perturbations.
In other words, the defender aims to build a defense mechanism that enables GNN classifiers to detect malicious traffic even in the presence of attackers executing topological manipulations within the network graph.
Moreover, the defender seeks to preserve or improve the NIDS's original detection accuracy on unperturbed input.

The defender has complete \textit{knowledge} of the underlying network infrastructure, as well as the internal configuration and operational details of the GNN-based NIDS, such as feature space, graph representation, and model parameters.
The defender has access to historical malicious traffic that previously triggered security alerts and was confirmed as part of verified security incidents.
However, the defender lacks prior knowledge of the adversary’s specific strategies or technical implementations, reflecting the practical challenge of defending against unknown attack variants.
Furthermore, the defender does not know exactly which hosts have been compromised by the attacker and thus cannot rely on knowledge of malicious nodes for their defense method.

The defender maintains full control over all machines within the private network.
In line with their role as system and network administrators, the defender's \textit{capabilities} include:
\begin{itemize}
\item Access historical network traffic data and corresponding labels (i.e., the training dataset);
\item Initiate legitimate communications and replace the endpoints involved in the transmission, thereby generating adversarial examples;
\item Expand the training set with the adversarial flows and modify the GNN classifier’s training process.
\end{itemize}
\section{Proposed Defense Strategy}
\label{sec:defensestrategy}
In this section, we describe the proposed defense framework to harden GNN-based NIDS against structural adversarial attacks.
As mentioned in Section~\ref{sec:backgroundrelatedwork}, our defense methodology is based on adversarial training.
Indeed, adversarial training is widely recognized as one of the most effective strategies for defending ML classifiers against adversarial attacks~\cite{szegedy2013intriguing, madry2017towards}.
Building on early work in other fields that demonstrates its effectiveness for GNN models~\cite{gosch2023adversarial}, we design and revisit this approach to enhance the robustness of GNN-based NIDS.

\begin{figure*}[t]
\centering
\includegraphics[width=\textwidth]{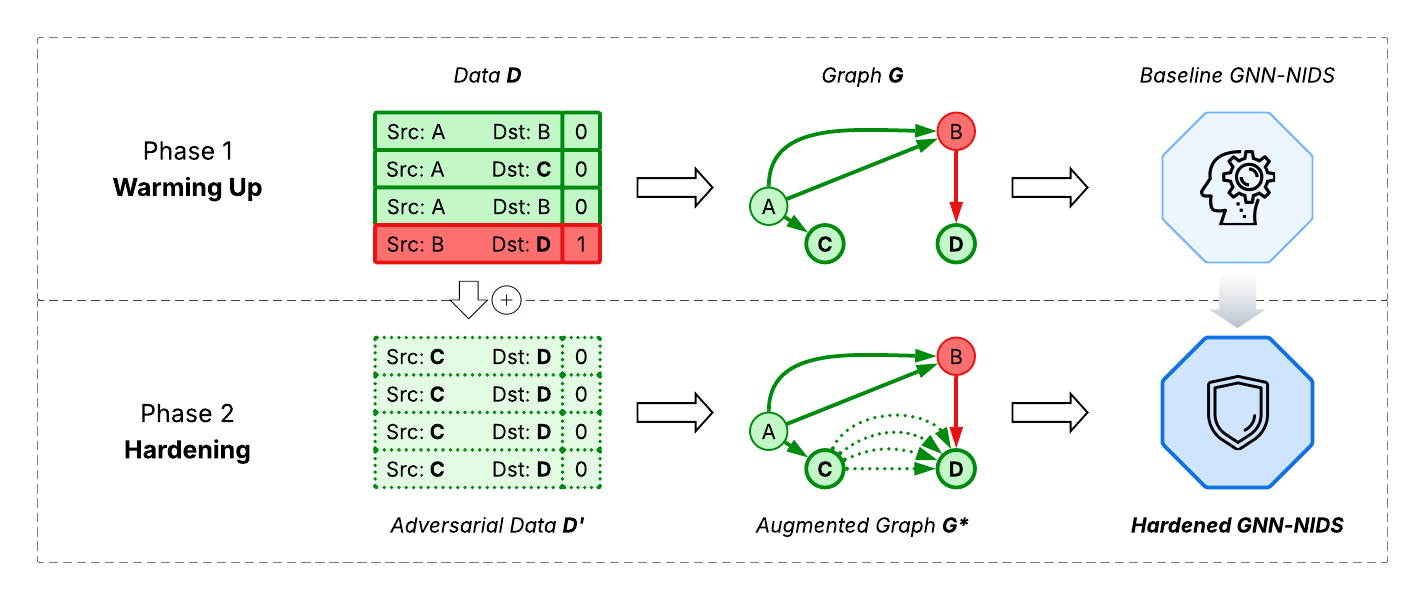}
\caption{Defense strategy to counter structural adversarial attacks against GNN-based NIDS.}
\label{fig:defensestrategy}
\end{figure*}

An overview of the proposed two-stage defense strategy is illustrated in Fig.~\ref{fig:defensestrategy}.
The considered example network contains four unique hosts indicated by the letters \textit{A}, \textit{B}, \textit{C}, and \textit{D}.
Each of the nodes \textit{C} and \textit{D} has only one incoming edge, giving it a \textit{degree} equal to $1$.
The GNN-based NIDS operates on flow graphs, where nodes represent network hosts and edges store the features of the respective netflows.
Even if our discussion focuses on flow graphs, the proposed methodology can be extended to other graph representations (e.g., line graphs).
As noted in Section~\ref{sec:backgroundrelatedwork}, features associated with edges can be directly translated to vertices.

As illustrated at the top of Fig.~\ref{fig:defensestrategy}, the first part of our methodology involves a \textit{warm-up phase}.
In this phase, the GNN classifier is trained exclusively on clean network data.
This part allows the model to learn the fundamental patterns of benign and malicious network communications.
The warm-up is essential for establishing stable decision boundaries that can later be refined using adversarial data~\cite{pang2020bag}.

The bottom of Fig.~\ref{fig:defensestrategy} depicts the \textit{hardening phase} of our strategy.
Unlike traditional adversarial training, our approach does not restart training from scratch; instead, it continues from the warm-up stage.
Since the nodes \textit{C} and \textit{D} are the least connected vertices in the graph, they are employed to replace both the original source and destination hosts in benign communications, generating adversarial structural flows.
Specifically, we iteratively sample subsets of increasing size from the benign netflows in the initial training set and perturb them by replacing their original endpoints with low-degree nodes, i.e., those having a degree below the first percentile of the degree distribution.
During this procedure, the training set is augmented with an increasing number of perturbed samples.
This process effectively injects new benign edges between reassigned nodes in the resulting graph, exposing the GNN detector to controlled structural noise.

In the following, a detailed explanation of the two phases.

\subsection{Warming Up}
\label{subsec:warmingup}
The \textit{warm-up phase} consists of an initial training procedure where the GNN-based NIDS is exposed to untouched network data.
This phase defines the classifier’s baseline capabilities on clean data before adversarial examples are injected into the training set.
The goal is to regularize the learning process by establishing decision boundaries that separate benign and malicious traffic.
Indeed, premature exposure to adversarial samples could destabilize gradient updates and hinder loss convergence~\cite{pang2020bag}.

We denote the original training set as $\mathcal{D}=\{(f_i,y_i)\}_{i=1}^{N}$, where $f_i$ is the $i$-th flow example and $y_i$ is its ground-truth label.
A netflow is considered \textit{benign} if $y_i=0$, and \textit{malicious} if $y_i=1$.
Each flow record $f_i$ contains both structural information (i.e., \textit{endpoints}) and traffic statistics (i.e., \textit{features}).
In particular, each netflow $f_i$ is defined by a tuple $(s_i,d_i,x_i)$, where $s_i$ denotes the \textit{source endpoint} (e.g., $109.80.75.71:36457$), $d_i$ the \textit{destination endpoint} (e.g., $147.32.84.188:8080$), and $x_i$ the \textit{feature vector} (e.g., \textit{duration}, \textit{bytes}, \textit{packets}).

Based on the training set $\mathcal{D}$, we build a flow graph $G=(V,E)$, where $V$ is the set of nodes corresponding to unique network endpoints, and $E$ is the set of edges corresponding to flow examples.
More specifically, each netflow $f_i$ generates an edge between nodes $s_i$ and $d_i$, attributed with the feature vector $x_i$ and the binary label $y_i$.

During warm-up, we train the GNN classifier on the clean flow graph $G$ for a fixed number of warm-up epochs.
The GNN parameters are optimized by minimizing a supervised loss function, enabling the model to encode stable representations of benign and malicious samples.

\subsection{Hardening}
\label{subsec:hardening}
After the warm-up phase, the GNN-based NIDS transitions into the \textit{hardening phase}, which constitutes the core of our defense approach.
In this stage, the training set is progressively expanded with manipulated flows that induce novel edges within the flow graph.
This augmentation prevents the classifier from being overwhelmed at the beginning of training.
Instead, the number of perturbed samples is gradually increased to enable the classifier to refine its decision boundaries.

To generate adversarial examples, we randomly sample a subset $\mathcal{B}_t$ from the benign flows $\mathcal{B}$.
The total amount of benign netflows sampled for manipulation is given by $\epsilon_t \cdot |\mathcal{B}_t|$, where $\epsilon_t$ is the perturbation intensity and $|\mathcal{B}_t|$ is the number of selected records.
Hence, the parameter $\epsilon_t$ controls the number of perturbed samples injected into the original training set.
The value of $\epsilon_t$ depends on epoch $t$ and is progressively increased every $T$ epochs during the hardening stage.
While the perturbation intensities can be chosen freely, we bound these values to maintain realistic network conditions and prevent excessive distortion of the original traffic patterns.
Indeed, we observe that large perturbations offer no advantage to the defender and may harm clean performance in non-adversarial scenarios~\cite{bostani2024effectiveness}.

For each sampled benign flow $f_i=(s_i,d_i,x_i) \in \mathcal{B}_t$, we generate its adversarial version $f'_i=(s'_i,d'_i,x_i) \in \mathcal{B}'_t$ by replacing both the source and destination endpoints $(s_i,d_i)$ with different endpoints $(s'_i,d'_i)$.
The feature vector $x_i$ remains unchanged, preserving original traffic attributes~\cite{pierazzi2020intriguing}.
We select as new endpoints $(s'_i,d'_i)$ the \textit{low-degree nodes} $V'$ of the graph obtained from the training set.
More specifically, we identify a pool of nodes whose connectivity falls below the first percentile of the total degree distribution of the original network graph and use it to perturb the structural information of the sampled netflows $\mathcal{B}_t$.
This operation simulates a practical \textit{edge injection attack}, establishing new communications from low-degree nodes to random vertices.
In fact, the edges in the network graph are relocated when we replace the endpoints of the pool of benign samples: (1) the edge between the initial source and destination nodes $(s_i,d_i)$ is dropped; (2) the feature vector $x_i$ is saved; (3) a new edge with the same features $x_i$ is set up between the replacement hosts $(s'_i,d'_i)$.

Finally, we combine the original examples $f_i$ with the newly generated flows $f'_i$ to form the augmented dataset $\mathcal{D}^\star$.
This expanded dataset $\mathcal{D}^\star$ is then converted into a graph $G^\star$, which serves to train the GNN using the supervised loss and model weights obtained at the end of the warm-up.
\section{Testbed}
\label{sec:experiments}
This section presents the experimental campaign conducted to validate the effectiveness of the proposed defense.
Section~\ref{subsec:datasets} describes the datasets employed, Section~\ref{subsec:detector} details the GNN-based detectors, and Section~\ref{subsec:implementation} outlines the implementation of our framework.
To promote reproducibility and transparency, we publicly release our code and experimental results\footnote{\url{https://github.com/dimgalli/defending-gnn-nids.git}}.

\subsection{Datasets}
\label{subsec:datasets}
We adopt two publicly available datasets: \textbf{CTU-13}~\cite{garcia2014empirical} and \textbf{TON-IoT}~\cite{alsaedi2020ton_iot}.
These two datasets are well-known benchmarks in the intrusion detection domain and are widely referenced in the literature~\cite{venturi2023arganids, galli2025evaluating}.
They are particularly suitable for our task, as they provide labeled netflows that can be readily transformed into graph-based representations~\cite{lo2022graphsage, venturi2023arganids}.
Together, both datasets encompass a broad spectrum of cyberattacks and structural patterns, making them ideal for evaluating the robustness of GNN-based NIDS.

On the one hand, CTU-13 comprises thirteen scenarios, each consisting of several hours of traffic from a medium-sized enterprise network.
Together with benign activity, each scenario includes malicious traffic generated by a specific botnet variant.
On the other hand, TON-IoT includes heterogeneous data collected from IoT environments, such as system logs, raw telemetry, and network traces.
In this work, we use only the flow records, which include multiple types of cyberattacks.

We preprocess both datasets following the methodology proposed in~\cite{venturi2021drelab}.
The resulting feature set includes source and destination endpoints, with more than $30$ statistical attributes, consistent with prior studies~\cite{venturi2023practical, galli2025evaluating}.
Benign flows $\mathcal{B}$ are merged into a single dataset, whereas malicious flows $\mathcal{M}$ associated with each threat $\alpha$ are grouped into dedicated subsets $\mathcal{M}_\alpha$.
We discard malicious variants with too few records.

For each threat $\alpha$, we construct a dataset $\mathcal{D}_\alpha$ by combining benign flows $\mathcal{B}$ with malicious samples $\mathcal{M}_\alpha$, using a benign-to-malicious ratio of $10:1$ to reflect the imbalance typically observed in operational networks.
Each dataset $\mathcal{D}_\alpha$ is divided into training and test sets with an $80:20$ split.

Table~\ref{tab:data} reports the resulting dataset statistics, i.e., the amount of malicious netflows together with the number of nodes and edges of the generated graphs.
Each row corresponds to a distinct malicious variant $\alpha$.

\begin{table}[t]
\centering
\caption{CTU-13 and TON-IoT datasets statistics.}
\label{tab:data}
\resizebox{\linewidth}{!}{%
\begin{tabular}{|c|c||ccc||ccc|}
\hline
\rowcolor[HTML]{EFEFEF}
\cellcolor[HTML]{EFEFEF} & \cellcolor[HTML]{EFEFEF} & \multicolumn{3}{c||}{\cellcolor[HTML]{EFEFEF}\textbf{Training}} & \multicolumn{3}{c|}{\cellcolor[HTML]{EFEFEF}\textbf{Testing}} \\
\cline{3-8}
\rowcolor[HTML]{EFEFEF}
\multirow{-2}{*}{\cellcolor[HTML]{EFEFEF}\textbf{Dataset}} & \multirow{-2}{*}{\cellcolor[HTML]{EFEFEF}\textbf{Threat}} & \multicolumn{1}{c|}{\cellcolor[HTML]{EFEFEF}\textit{\#Mal}} & \multicolumn{1}{c|}{\cellcolor[HTML]{EFEFEF}\textit{\#Nodes}} & \multicolumn{1}{c||}{\cellcolor[HTML]{EFEFEF}\textit{\#Edges}} & \multicolumn{1}{c|}{\cellcolor[HTML]{EFEFEF}\textit{\#Mal}} & \multicolumn{1}{c|}{\cellcolor[HTML]{EFEFEF}\textit{\#Nodes}} & \multicolumn{1}{c|}{\cellcolor[HTML]{EFEFEF}\textit{\#Edges}} \\
\hline
\hline
& \textit{Neris} & \multicolumn{1}{c|}{$64\,076$} & \multicolumn{1}{c|}{$540\,679$} & \multicolumn{1}{c||}{$1\,409\,672$} & \multicolumn{1}{c|}{$16\,019$} & \multicolumn{1}{c|}{$167\,881$} & \multicolumn{1}{c|}{$352\,418$} \\
\cline{2-8}
& \textit{Rbot} & \multicolumn{1}{c|}{$22\,007$} & \multicolumn{1}{c|}{$228\,374$} & \multicolumn{1}{c||}{$484\,158$} & \multicolumn{1}{c|}{$5\,502$} & \multicolumn{1}{c|}{$68\,611$} & \multicolumn{1}{c|}{$121\,040$} \\
\cline{2-8}
& \textit{Virut} & \multicolumn{1}{c|}{$25\,875$} & \multicolumn{1}{c|}{$252\,747$} & \multicolumn{1}{c||}{$569\,254$} & \multicolumn{1}{c|}{$6\,469$} & \multicolumn{1}{c|}{$75\,734$} & \multicolumn{1}{c|}{$142\,314$} \\
\cline{2-8}
& \textit{Menti} & \multicolumn{1}{c|}{$2\,260$} & \multicolumn{1}{c|}{$29\,327$} & \multicolumn{1}{c||}{$49\,720$} & \multicolumn{1}{c|}{$565$} & \multicolumn{1}{c|}{$8\,339$} & \multicolumn{1}{c|}{$12\,430$} \\
\cline{2-8}
\multicolumn{1}{|c|}{\multirow{-5}{*}{\rotatebox[origin=c]{90}{\textbf{CTU-13}}}} & \textit{Murlo} & \multicolumn{1}{c|}{$885$} & \multicolumn{1}{c|}{$12\,341$} & \multicolumn{1}{c||}{$19\,464$} & \multicolumn{1}{c|}{$221$} & \multicolumn{1}{c|}{$3\,392$} & \multicolumn{1}{c|}{$4\,868$} \\
\hline
\hline
& \textit{Bkdr} & \multicolumn{1}{c|}{$10\,522$} & \multicolumn{1}{c|}{$27\,449$} & \multicolumn{1}{c||}{$231\,496$} & \multicolumn{1}{c|}{$2\,631$} & \multicolumn{1}{c|}{$9\,173$} & \multicolumn{1}{c|}{$57\,874$} \\
\cline{2-8}
& \textit{DDoS} & \multicolumn{1}{c|}{$10\,522$} & \multicolumn{1}{c|}{$36\,683$} & \multicolumn{1}{c||}{$231\,496$} & \multicolumn{1}{c|}{$2\,631$} & \multicolumn{1}{c|}{$11\,209$} & \multicolumn{1}{c|}{$57\,874$} \\
\cline{2-8}
& \textit{DoS} & \multicolumn{1}{c|}{$10\,522$} & \multicolumn{1}{c|}{$28\,188$} & \multicolumn{1}{c||}{$231\,496$} & \multicolumn{1}{c|}{$2\,631$} & \multicolumn{1}{c|}{$9\,362$} & \multicolumn{1}{c|}{$57\,874$} \\
\cline{2-8}
& \textit{Inj} & \multicolumn{1}{c|}{$10\,522$} & \multicolumn{1}{c|}{$36\,265$} & \multicolumn{1}{c||}{$231\,496$} & \multicolumn{1}{c|}{$2\,631$} & \multicolumn{1}{c|}{$11\,169$} & \multicolumn{1}{c|}{$57\,874$} \\
\cline{2-8}
& \textit{Pswd} & \multicolumn{1}{c|}{$10\,522$} & \multicolumn{1}{c|}{$36\,356$} & \multicolumn{1}{c||}{$231\,496$} & \multicolumn{1}{c|}{$2\,631$} & \multicolumn{1}{c|}{$11\,191$} & \multicolumn{1}{c|}{$57\,874$} \\
\cline{2-8}
& \textit{Rans} & \multicolumn{1}{c|}{$10\,522$} & \multicolumn{1}{c|}{$28\,608$} & \multicolumn{1}{c||}{$231\,496$} & \multicolumn{1}{c|}{$2\,631$} & \multicolumn{1}{c|}{$9\,680$} & \multicolumn{1}{c|}{$57\,874$} \\
\cline{2-8}
& \textit{Scan} & \multicolumn{1}{c|}{$10\,522$} & \multicolumn{1}{c|}{$32\,089$} & \multicolumn{1}{c||}{$231\,496$} & \multicolumn{1}{c|}{$2\,631$} & \multicolumn{1}{c|}{$10\,753$} & \multicolumn{1}{c|}{$57\,874$} \\
\cline{2-8}
\multicolumn{1}{|c|}{\multirow{-8}{*}{\rotatebox[origin=c]{90}{\textbf{ToN-IoT}}}} & \textit{XSS} & \multicolumn{1}{c|}{$10\,522$} & \multicolumn{1}{c|}{$33\,874$} & \multicolumn{1}{c||}{$231\,496$} & \multicolumn{1}{c|}{$2\,631$} & \multicolumn{1}{c|}{$10\,960$} & \multicolumn{1}{c|}{$57\,874$} \\
\hline
\end{tabular}%
}
\end{table}

\subsection{Detector}
\label{subsec:detector}
In our experimental campaign, we target a NIDS based on the \textbf{E-GraphSAGE}~\cite{lo2022graphsage} architecture. 
It is an extension of the popular GraphSAGE model~\cite{hamilton2017inductive} that performs edge classification for network intrusion detection.

E-GraphSAGE converts network traffic into a flow graph, where each unique endpoint is represented as a node, and each netflow corresponds to an edge linking the endpoints included in it.
Each edge is attributed with the record’s statistical features (e.g., duration, byte count, packet size).

The model processes the input through multiple graph convolutional layers to generate low-dimensional vectors for each edge.
These embeddings encode both the edge’s features and information from its local neighborhood.
Importantly, the inductive nature of E-GraphSAGE allows the generation of embeddings for previously unseen nodes at inference time.
This is a key property for a GNN-based NIDS operating in dynamic and realistic network environments.

Following best practices~\cite{apruzzese2023sok}, the detector is organized as an ensemble of binary classifiers, each implemented as a specialized E-GraphSAGE model trained to detect a specific malicious class $\alpha$.
All classifiers are configured with the same hyperparameters as in the original work~\cite{lo2022graphsage} to ensure consistency.

\subsection{Framework Implementation}
\label{subsec:implementation}
We now describe in detail the implementation of our framework (Section~\ref{sec:defensestrategy}).
All experiments were implemented in \textit{Python}, using \textit{pandas} and \textit{scikit-learn} for data preprocessing and graph generation, and \textit{DGL} with \textit{PyTorch Geometric} for GNN training and testing.

As outlined in Section~\ref{subsec:datasets}, each dataset $\mathcal{D}_\alpha$ is split into training and test sets using an $80:20$ ratio while preserving a $10:1$ benign-to-malicious class distribution.
First, we obtain clean flow graphs from the training data and use them to train a \textit{baseline} NIDS composed of an ensemble of E-GraphSAGE classifiers, each specialized in detecting a specific malicious variant $\alpha$ (Section~\ref{subsec:detector}).

We also train a hardened version of each classifier for comparison.
As discussed in Section~\ref{subsec:warmingup}, each E-GraphSAGE classifier undergoes a warm-up phase of $100$ epochs on the unperturbed flow graph to establish stable decision boundaries.
The learning process then continues for an additional $100$ epochs on adversarially augmented data, starting from the loss value and model weights obtained at the end of the warm-up.
As explained in Section~\ref{subsec:hardening}, we collect and perturb progressively larger sets of benign flows.
The number of benign samples selected for manipulation at epoch $t$ is controlled by the perturbation intensity $\epsilon_t$, which starts at $1$ and increases by $1$ every $T=20$ epochs up to a maximum of $5$.
We cap $\epsilon_t$ at $5$ to ensure realistic structural conditions and avoid excessive distortion of decision boundaries that could degrade clean performance.

At each step, benign records $\mathcal{B}_t$ are modified by replacing their source and destination endpoints with low-degree nodes.
The perturbed samples $\mathcal{B}'_t$ are injected into the training set to generate the adversarially augmented dataset $\mathcal{D}^\star$, which is transformed into the expanded graph $G^\star$ that can be used to harden E-GraphSAGE models.
\section{Experimental Results}
\label{sec:results}
This section presents the results of our experimental campaign, which aims to prove two key points:
(1) our defense yields a GNN-based NIDS whose performance on clean network graphs remains comparable or superior to state-of-the-art models
and
(2) the proposed method significantly enhances resilience against structural adversarial attacks.

We compare the detection performance of the baseline and hardened models across two different evaluation scenarios: \textit{clean performance} on unperturbed data (Section~\ref{subsec:cleanperformance}) and \textit{adversarial robustness} against structural attacks (Section~\ref{subsec:adversarialrobustness}).

In the first scenario, we evaluate the detection accuracy of both classifiers on clean inputs.
We target both classifiers with test flow graphs that do not involve any structural manipulation.
These experiments verify that both classifiers correctly detect the malicious variants they were trained on without worsening clean performance under standard operating conditions.
This constraint is critical, as a NIDS that performs poorly on known attacks is unsuitable for practical deployment in real-world networks.

In the second scenario, we measure the adversarial robustness of both detectors against realistic structural attacks.
We employ the \CtoXB, \CtoXM, and \textbf{add node} structural attacks proposed in~\cite{venturi2024problem} for this evaluation.
These attacks comply with real-world constraints and have demonstrated high effectiveness against state-of-the-art GNN detectors.
Hence, they serve as a representative benchmark for evaluating robustness in adversarial settings.

For evaluating the classifiers on clean data, we report performance using standard ML metrics from the literature, i.e., \textit{F1-score}, \textit{Precision}, and \textit{Recall} (\textit{Detection Rate}).
These metrics are defined as follows:
\begin{equation}
F1=2 \times \frac{Precision \times Recall}{Precision + Recall}
\end{equation}
\begin{equation}
Precision=\frac{TP}{TP+FP}
\end{equation}
\begin{equation}
Recall=\frac{TP}{TP+FN}
\end{equation}
where \textit{TP}, \textit{FP}, and \textit{FN} denote \textit{true positives}, \textit{false positives}, and \textit{false negatives}, respectively.
For the evaluation of the adversarial robustness, we use the \textit{Detection Rate (DR)}.
In the paper, we consider a positive detection as a malicious sample.

\subsection{Clean Performance}
\label{subsec:cleanperformance}
We first evaluate the performance of the baseline and hardened E-GraphSAGE classifiers on clean flow graphs.

We report the results in Table~\ref{tab:cleanperformance}: each row compares the scores of both classifiers for a specific malicious activity, while the columns correspond to the performance metrics.
For each class, we highlight the best scores in bold.
For each dataset, the bottom row presents the mean ($\mu$) and standard deviation ($\sigma$) of the metrics values across all models.

\begin{table*}[t]
\centering
\caption{Clean performance scores of E-GraphSAGE on CTU-13 and TON-IoT datasets.}
\label{tab:cleanperformance}
\resizebox{0.75\textwidth}{!}{%
\begin{tabular}{|c|c||cc||cc||cc|}
\hline
\rowcolor[HTML]{EFEFEF}
\cellcolor[HTML]{EFEFEF} & \cellcolor[HTML]{EFEFEF} & \multicolumn{2}{c||}{\cellcolor[HTML]{EFEFEF}\textbf{\textit{F1-score}}} & \multicolumn{2}{c||}{\cellcolor[HTML]{EFEFEF}\textbf{\textit{Precision}}} & \multicolumn{2}{c|}{\cellcolor[HTML]{EFEFEF}\textbf{\textit{Recall}}} \\
\cline{3-8}
\rowcolor[HTML]{EFEFEF}
\multirow{-2}{*}{\cellcolor[HTML]{EFEFEF}\textbf{Dataset}} & \multirow{-2}{*}{\cellcolor[HTML]{EFEFEF}\textbf{Threat}} & \multicolumn{1}{c|}{\cellcolor[HTML]{EFEFEF}\textit{Baseline}} & \multicolumn{1}{c||}{\cellcolor[HTML]{EFEFEF}\textit{Hardened}} & \multicolumn{1}{c|}{\cellcolor[HTML]{EFEFEF}\textit{Baseline}} & \multicolumn{1}{c||}{\cellcolor[HTML]{EFEFEF}\textit{Hardened}} & \multicolumn{1}{c|}{\cellcolor[HTML]{EFEFEF}\textit{Baseline}} & \multicolumn{1}{c|}{\cellcolor[HTML]{EFEFEF}\textit{Hardened}} \\
\hline
\hline
\multicolumn{1}{|c|}{} & \textit{Neris} & \multicolumn{1}{c|}{$0.880$} & \bm{$0.891$} & \multicolumn{1}{c|}{$0.948$} & \bm{$0.954$} & \multicolumn{1}{c|}{$0.821$} & \bm{$0.836$} \\
\cline{2-8}
\multicolumn{1}{|c|}{} & \textit{Rbot} & \multicolumn{1}{c|}{$0.989$} & \bm{$0.989$} & \multicolumn{1}{c|}{$0.987$} & \bm{$0.988$} & \multicolumn{1}{c|}{\bm{$0.991$}} & $0.990$ \\
\cline{2-8}
\multicolumn{1}{|c|}{} & \textit{Virut} & \multicolumn{1}{c|}{$0.945$} & \bm{$0.945$} & \multicolumn{1}{c|}{\bm{$0.979$}} & $0.947$ & \multicolumn{1}{c|}{$0.913$} & \bm{$0.944$} \\
\cline{2-8}
\multicolumn{1}{|c|}{} & \textit{Menti} & \multicolumn{1}{c|}{$0.996$} & \bm{$0.996$} & \multicolumn{1}{c|}{$0.993$} & \bm{$0.993$} & \multicolumn{1}{c|}{$1.000$} & \bm{$1.000$} \\
\cline{2-8}
\multicolumn{1}{|c|}{} & \textit{Murlo} & \multicolumn{1}{c|}{$0.984$} & \bm{$0.991$} & \multicolumn{1}{c|}{\bm{$0.991$}} & $0.987$ & \multicolumn{1}{c|}{$0.977$} & \bm{$0.995$} \\
\cline{2-8}
\multicolumn{1}{|c|}{\multirow{-6}{*}{\rotatebox[origin=c]{90}{\textbf{CTU-13}}}} & \cellcolor[HTML]{EFEFEF}\textit{\begin{tabular}[c]{@{}c@{}}$\mu$ \\ ($\sigma$)\end{tabular}} & \multicolumn{1}{c|}{\cellcolor[HTML]{EFEFEF}\begin{tabular}[c]{@{}c@{}}$0.959$ \\ ($0.048$)\end{tabular}} & \multicolumn{1}{c||}{\cellcolor[HTML]{EFEFEF}\begin{tabular}[c]{@{}c@{}}\bm{$0.962$} \\ ($0.045$)\end{tabular}} & \multicolumn{1}{c|}{\cellcolor[HTML]{EFEFEF}\begin{tabular}[c]{@{}c@{}}\bm{$0.980$} \\ ($0.018$)\end{tabular}} & \multicolumn{1}{c||}{\cellcolor[HTML]{EFEFEF}\begin{tabular}[c]{@{}c@{}}$0.974$ \\ ($0.022$)\end{tabular}} & \multicolumn{1}{c|}{\cellcolor[HTML]{EFEFEF}\begin{tabular}[c]{@{}c@{}}$0.940$ \\ ($0.075$)\end{tabular}} & \multicolumn{1}{c|}{\cellcolor[HTML]{EFEFEF}\begin{tabular}[c]{@{}c@{}}\bm{$0.953$} \\ ($0.069$)\end{tabular}} \\
\hline
\hline
\multicolumn{1}{|c|}{} & \textit{Backdoor} & \multicolumn{1}{c|}{$0.997$} & \bm{$0.997$} & \multicolumn{1}{c|}{\bm{$0.998$}} & $0.996$ & \multicolumn{1}{c|}{$0.995$} & \bm{$0.998$} \\
\cline{2-8}
\multicolumn{1}{|c|}{} & \textit{DDoS} & \multicolumn{1}{c|}{\bm{$0.992$}} & $0.991$ & \multicolumn{1}{c|}{\bm{$0.993$}} & $0.990$ & \multicolumn{1}{c|}{$0.990$} & \bm{$0.991$} \\
\cline{2-8}
\multicolumn{1}{|c|}{} & \textit{DoS} & \multicolumn{1}{c|}{$0.993$} & \bm{$0.994$} & \multicolumn{1}{c|}{$0.995$} & \bm{$0.995$} & \multicolumn{1}{c|}{$0.992$} & \bm{$0.994$} \\
\cline{2-8}
\multicolumn{1}{|c|}{} & \textit{Injection} & \multicolumn{1}{c|}{$0.991$} & \bm{$0.994$} & \multicolumn{1}{c|}{\bm{$0.991$}} & $0.989$ & \multicolumn{1}{c|}{$0.992$} & \bm{$1.000$} \\
\cline{2-8}
\multicolumn{1}{|c|}{} & \textit{Password} & \multicolumn{1}{c|}{\bm{$0.994$}} & $0.992$ & \multicolumn{1}{c|}{\bm{$0.995$}} & $0.987$ & \multicolumn{1}{c|}{$0.994$} & \bm{$0.996$} \\
\cline{2-8}
\multicolumn{1}{|c|}{} & \textit{Ransomware} & \multicolumn{1}{c|}{$0.998$} & \bm{$0.999$} & \multicolumn{1}{c|}{$0.997$} & \bm{$0.998$} & \multicolumn{1}{c|}{$1.000$} & \bm{$1.000$} \\
\cline{2-8}
\multicolumn{1}{|c|}{} & \textit{Scanning} & \multicolumn{1}{c|}{$0.996$} & \bm{$0.996$} & \multicolumn{1}{c|}{$0.997$} & \bm{$0.997$} & \multicolumn{1}{c|}{$0.995$} & \bm{$0.995$} \\
\cline{2-8}
\multicolumn{1}{|c|}{} & \textit{XSS} & \multicolumn{1}{c|}{$0.987$} & \bm{$0.995$} & \multicolumn{1}{c|}{\bm{$0.993$}} & $0.990$ & \multicolumn{1}{c|}{$0.981$} & \bm{$1.000$} \\
\cline{2-8}
\multicolumn{1}{|c|}{\multirow{-9}{*}{\rotatebox[origin=c]{90}{\textbf{TON-IoT}}}} & \cellcolor[HTML]{EFEFEF}\textit{\begin{tabular}[c]{@{}c@{}}$\mu$ \\ ($\sigma$)\end{tabular}} & \multicolumn{1}{c|}{\cellcolor[HTML]{EFEFEF}\begin{tabular}[c]{@{}c@{}}$0.994$ \\ ($0.004$)\end{tabular}} & \multicolumn{1}{c||}{\cellcolor[HTML]{EFEFEF}\begin{tabular}[c]{@{}c@{}}\bm{$0.995$} \\ ($0.003$)\end{tabular}} & \multicolumn{1}{c|}{\cellcolor[HTML]{EFEFEF}\begin{tabular}[c]{@{}c@{}}\bm{$0.995$} \\ ($0.003$)\end{tabular}} & \multicolumn{1}{c||}{\cellcolor[HTML]{EFEFEF}\begin{tabular}[c]{@{}c@{}}$0.993$ \\ ($0.004$)\end{tabular}} & \multicolumn{1}{c|}{\cellcolor[HTML]{EFEFEF}\begin{tabular}[c]{@{}c@{}}$0.992$ \\ ($0.005$)\end{tabular}} & \multicolumn{1}{c|}{\cellcolor[HTML]{EFEFEF}\begin{tabular}[c]{@{}c@{}}\bm{$0.997$} \\ ($0.003$)\end{tabular}} \\
\hline
\end{tabular}%
}
\end{table*}

The results show that nearly all values exceed $0.9$, confirming the strong detection capabilities of E-GraphSAGE.
The only isolated case concerns the \textit{Neris} botnet of the \textit{CTU-13} dataset, where the baseline and hardened detectors yield slightly lower \textit{F1-scores} of $0.880$ and $0.891$, respectively.
In most cases, the hardened classifiers outperform their baseline counterparts, achieving average \textit{F1-scores} of $0.962$ versus $0.959$ on \textit{CTU-13}, and $0.995$ versus $0.994$ on \textit{TON-IoT}.

Overall, E-GraphSAGE models demonstrate high performance consistent with state-of-the-art GNN-based NIDS~\cite{lo2022graphsage,venturi2023practical}.
These findings endorse that our defense framework maintains good performance under clean conditions, slightly improving \textit{F1-scores}, even when \textit{Precision} experiences minor decreases.
The strong performance of E-GraphSAGE on unperturbed data indicates that the hardened NIDS is suitable for real-world deployment and motivates its further evaluation under adversarial conditions (Section~\ref{subsec:adversarialrobustness}).

\subsection{Adversarial Robustness}
\label{subsec:adversarialrobustness}
We now evaluate the resilience of the baseline and hardened E-GraphSAGE classifiers against adversarial attacks involving structural perturbations.
Building on previous works~\cite{venturi2024problem}, attackers can perturb the test flow graph in three distinct ways:
\begin{itemize}
\item \textit{\CtoXB~attacks}: compromised hosts establish benign communication to random targets;
\item \textit{\CtoXM~attacks}: controlled machines initiate malicious connections to random targets;
\item \textit{\textbf{Add node}~attacks}: new, previously unseen endpoints generate benign flows.
\end{itemize}
This set of attacks represents the main perturbation strategies available to a capable attacker.
The goal is to determine whether our adversarial training approach enhances model robustness under such conditions.

\begin{figure*}[htbp]
\centering
\begin{subfigure}[htbp]{\textwidth}
\centering
\includegraphics[width=\textwidth]{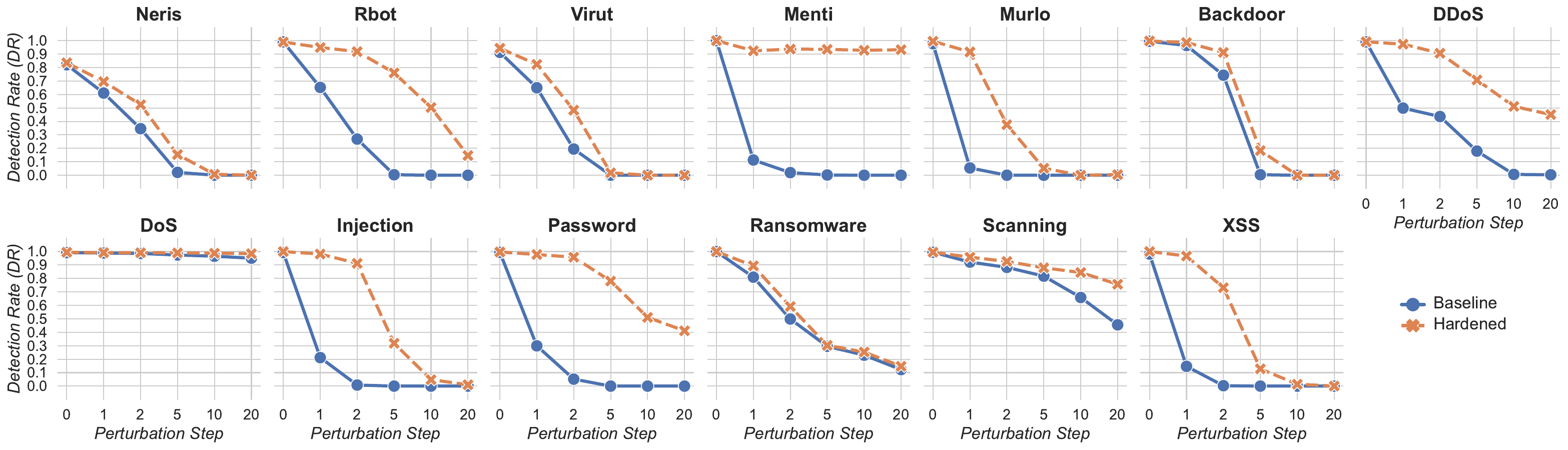}
\caption{\CtoXB~attacks.}
\label{fig:c2x_b}
\end{subfigure}
\begin{subfigure}[htbp]{\textwidth}
\centering
\includegraphics[width=\textwidth]{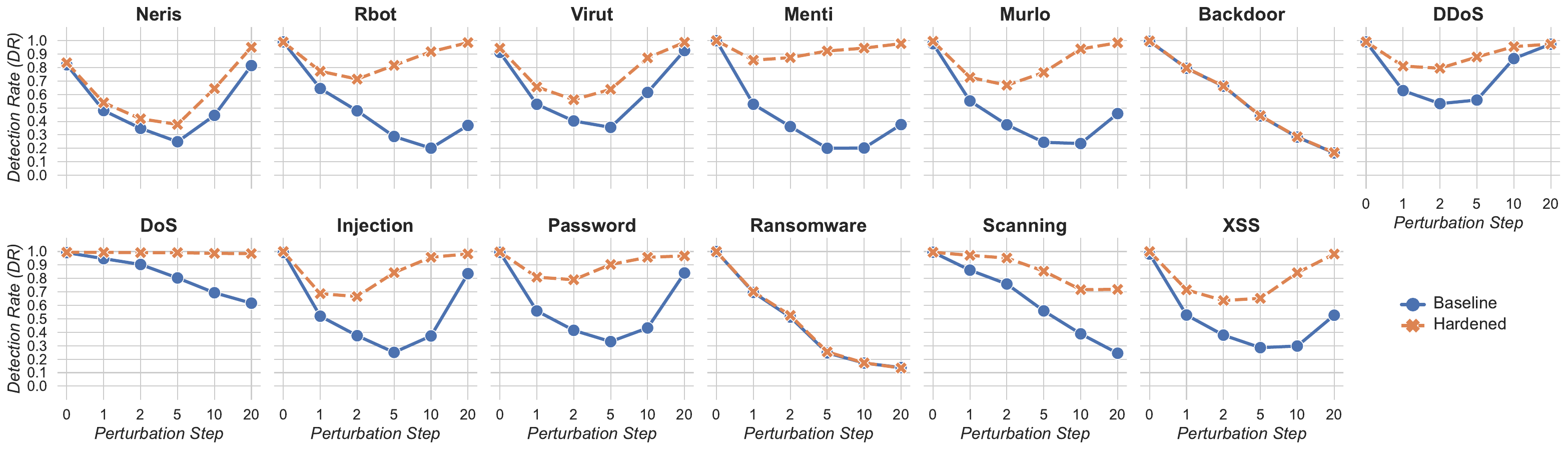}
\caption{\CtoXM~attacks.}
\label{fig:c2x_m}
\end{subfigure}
\begin{subfigure}[htbp]{\textwidth}
\centering
\includegraphics[width=\textwidth]{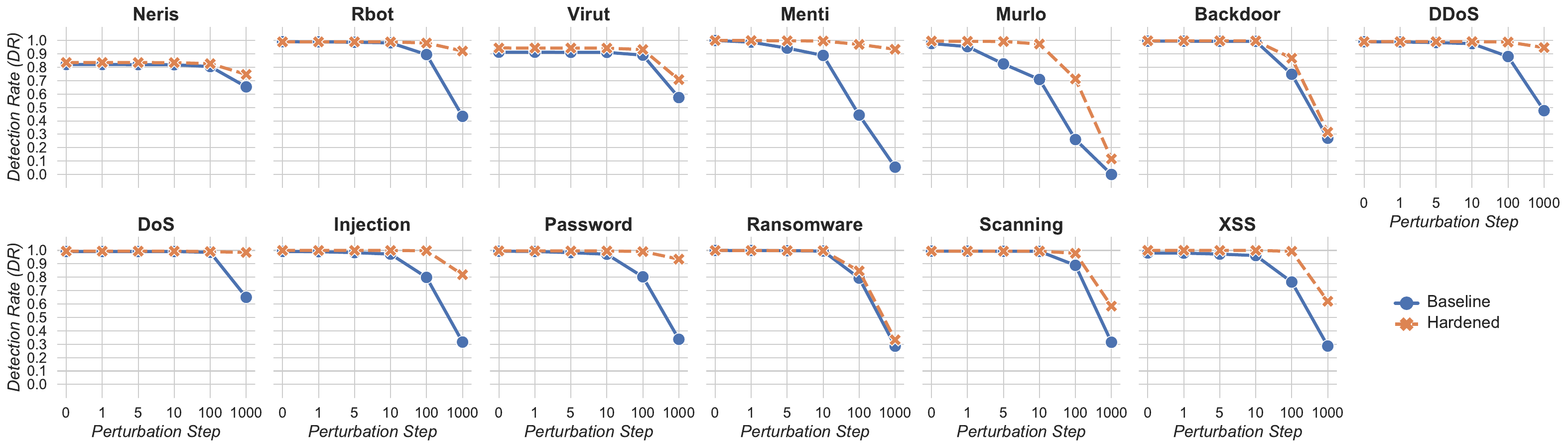}
\caption{\textbf{Add node}~attacks.}
\label{fig:add_node}
\end{subfigure}    
\caption{Adversarial robustness scores of E-GraphSAGE against structural attacks.}
\label{fig:adversarialrobustness}
\end{figure*}

The results are shown in Fig.~\ref{fig:adversarialrobustness}, with separate subfigures for each attack type: Fig.~\ref{fig:c2x_b} for \textit{\CtoXB~attacks}, Fig.~\ref{fig:c2x_m} for \textit{\CtoXM~attacks}, and Fig.~\ref{fig:add_node} for \textit{\textbf{add node}~attacks}.
Each plot reports the \textit{Detection Rate (DR)} (y-axis) of both classifiers as the \textit{perturbation step} (x-axis) increases.
For \CtoXB~and \CtoXM~attacks, the perturbation step indicates the number of perturbed flows injected into the test set (i.e., added edges from compromised nodes to random ones), with values of $0$, $1$, $2$, $5$, $10$, and $20$.
For \textbf{add node} attacks, it represents the number of new nodes injected into the resulting test graph, with values of $0$, $1$, $5$, $10$, $100$, and $1000$.
We have the same model performance on clean data (Section~\ref{subsec:cleanperformance}) with a perturbation step of $0$.
Each curve in the plots represents the robustness trend (in terms of \textit{DR}) for a specific detector variant.

We observe that the baseline classifiers, trained on clean data, are highly vulnerable to all types of attacks.
This behavior aligns with the findings reported in~\cite{venturi2024problem}, thereby empirically motivating our study and validating our experimental setup.

Across nearly all scenarios, \CtoXB~attacks prove to be the most damaging.
This observation supports our decision to employ manipulated benign netflows during adversarial training, which enhances the model’s ability to recognize malicious patterns hidden within seemingly legitimate connections.
By contrast, \CtoXM~attacks exhibit a different trend: the \textit{DR} curve does not decline monotonically but often increases beyond a certain perturbation threshold.
As noted in~\cite{venturi2024problem}, this occurs because large numbers of malicious samples create topological patterns that are easier for the detector to identify.
Conversely, \textbf{add node}~attacks are the least effective.
Indeed, they affect E-GraphSAGE performance only when attackers inject more than $100$ new nodes within the graph.

Most importantly, our adversarial training framework yields substantial gains in robustness.
This is evident from the slower decline in \textit{DR} of the hardened detectors compared to the baseline models, particularly for the \textit{Rbot}, \textit{Menti}, \textit{Murlo}, \textit{DDoS}, and \textit{Password} variants.
In particular, for the \textit{Menti} botnet under \CtoXB~attacks, the average \textit{DR} across all perturbation steps rises from $0.027$ of the baseline classifier to $0.926$ of the hardened one, corresponding to an absolute improvement of nearly $90\%$.

Even under high-intensity perturbations, the hardened classifiers consistently outperform their baseline counterparts, validating our hypothesis that adversarial training with low-intensity perturbations effectively enhances GNN resilience to more severe structural attacks.
However, none of the hardened classifiers achieve perfect detection, highlighting that there is still room for further improvement.
Overall, augmenting the training data with structurally perturbed benign netflows proves to be a simple yet powerful strategy for strengthen GNN detectors against structural adversarial attacks.
\section{Conclusion}
\label{sec:conclusion}
GNN-based NIDS represent a significant advancement in cybersecurity, as they leverage graph-structured representations of network traffic to capture complex relational patterns that traditional approaches often miss.
However, their dependence on graph topology also makes them uniquely vulnerable to structural adversarial attacks, a realistic threat that can undermine their reliability in practical deployments.
Existing countermeasures often fall short due to unrealistic assumptions or prohibitive computational costs, leaving a critical gap in the defensive landscape.

In this paper, we introduced and validated a novel defense based on structural adversarial training specifically designed to harden GNN-based NIDS against such attacks.
The core innovation of our framework lies in a lightweight yet effective mechanism for generating adversarial samples by selectively perturbing the source and destination hosts of benign netflows, thereby emulating realistic edge injection attacks.
Our experimental campaign on CTU-13 and TON-IoT datasets, using E-GraphSAGE as the base GNN classifier, yielded two central findings:
(1) our hardened detectors exhibit significantly improved resilience against diverse structural adversarial attacks
and
(2) this increased robustness is obtained without compromising (and in many scenarios even slightly improving) their performance on unperturbed inputs.

These results have significant implications for the deployment of GNN-based NIDS in operational environments.
Our method offers an efficient and practical defense strategy that does not rely on pre-existing adversarial data or costly perturbation-generation procedures.
By focusing on low-degree node replacement for benign flow perturbations, we demonstrate that adversarial training can meaningfully strengthen the resilience of GNN detectors against realistic structural threats.

\section*{Acknowledgment}
This work has been supported by the project C4SI funded by the Emilia-Romagna region within the PR-FESR 2021-2027 (CUP E67G22000630003).
This work has been supported by the European Union’s Horizon 2020 Research and Innovation Program under the project FALCON (Grant Agreement No. 101121281).

\bibliographystyle{IEEEtran}
\bibliography{references}

\end{document}